\documentclass[fleqn,usenatbib]{mnras}

\usepackage{newtxtext,newtxmath}
\usepackage[T1]{fontenc}

\DeclareRobustCommand{\VAN}[3]{#2}
\let\VANthebibliography\thebibliography
\def\thebibliography{\DeclareRobustCommand{\VAN}[3]{##3}\VANthebibliography}

\usepackage{graphicx}	% Including figure files
\usepackage{amsmath}	% Advanced maths commands

\usepackage{color}

\definecolor{ineke}{rgb}{0.5, 0., 1.0}
\definecolor{cosima}{rgb}{0.1, 0.5, 0.2}
\definecolor{sami}{rgb}{1, 0., 0.}

\newlength\mylength
\title[Synthetic Observables]{Effect of Numerical Resolution on Synthetic Observables of Simulated Coronal Loops}

\author[C. A. Breu et al.]{
C. A. Breu,$^{1}$\thanks{E-mail: cab42@st-andrews.ac.uk (University of St. Andrews)}
I. De Moortel,$^{1,2}$
H. Peter,$^{3,4}$
S.K. Solanki$^{3}$
\\
$^{1}$ School of Mathematics and Statistics, University of St Andrews, St Andrews, Fife KY16 9SS, UK\\
$^{2}$Rosseland Centre for Solar Physics, University of Oslo, PO Box 1029 Blindern, NO-0315 Oslo, Norway\\
$^{3}$Max Planck Institute for Solar System Research,  Justus-von-Liebig-Weg 3, 37077 G\"ottingen, Germany\\
$^{4}$ Institut f\"ur Sonnenphysik (KIS) 79110 Freiburg, Germany\\
}

\date{Accepted XXX. Received YYY; in original form ZZZ}

\pubyear{2025}

\begin{document}
\label{firstpage}
\pagerange{\pageref{firstpage}--\pageref{lastpage}}
\maketitle

\begin{abstract}
Increasingly realistic simulations of the corona are used to predict synthetic observables for instruments onboard both existing and upcoming heliophysics space missions. Synthetic observables play an important role in constraining coronal heating theories.
Choosing the spatial resolution of numerical simulations involves a trade-off between accuracy and computational cost. Since the numerical resolution not only affects the scale of structures that can be resolved, but also thermodynamic quantities such as the average coronal density, it is important to quantify the effect on synthesized observables.
Using 3D radiative MHD simulations of coronal loops at three different grid spacings, from 60 km down to 12 km, we find that changes in numerical resolution lead to differences in thermodynamic quantities and stratification as well as dynamic behaviour. Higher grid resolution results in a more complex and dynamic atmosphere. The resolution affects the emission intensity as well as the velocity distribution, thereby affecting synthetic spectra derived from the simulation. The distribution of synthetic coronal loop strand sizes changes as more fine-scale structure is resolved. A number of parameters, however, seem to start to saturate from our chosen medium grid resolution on.
Our study shows that while choosing a sufficiently high resolution matters when comparing forward-modelled observables with data from current and future space missions, for most purposes not much is gained by further increasing the resolution beyond a grid spacing of 24 km, which seems to be adequate for reproducing bulk loop properties and forward-modelled emission, representing a good trade-off between accuracy and computational resource. 
\end{abstract}


%%%%%%%%%%%%%%%%%%%%%%%%%%%%%%%%%%%%%%%%%%%%%%%%%%

%%%%%%%%%%%%%%%%% BODY OF PAPER %%%%%%%%%%%%%%%%%%

\section{Introduction}

Numerical models of coronal structures in the form of 3D magnetohydrodynamics (MHD) simulations make it possible to synthesize emission as it would be observed by space-based instruments and are an important tool to make predictions for ongoing and future facilities. It is in the nature of numerical simulations, however, that simplifying assumptions have to be made about the system being modelled. A critical challenge for numerical simulations of the solar corona is the large range of length scales involved in the coronal heating problem and the limitations placed on the spatial numerical resolution by the available computational power. While coronal structures such as plasma loops and prominences can have an extent of hundreds of Megameters (at least in one spatial dimension), expected widths of current sheets, the dissipative structures in the solar corona, are on the order of meters \citep{2015_Peter}. By necessity, numerical models are either limited to a small region such as a single current sheet, treat coronal structures in one dimension \citep{Bradshaw_2013} or attempt to model the behaviour of large structures at the expense of accurately incorporating the microphysics responsible for energy dissipation.  In numerical MHD simulations of the solar corona, heating occurs mostly at the smallest resolved spatial scales, while the true dissipation scale would be much smaller \citep{2015_Peter,2017_Rempel}.\\
Many existing simulations of active regions and other large-scale structures have a coarse grid spacing of 100 km or above \citep{2015_Peter}. For example, the simulation of a flare-producing active region by \citet{2023_Chen} has grid spacings of 192 and 64 km in the horizontal and vertical direction, while the simulation by \citet{2023_Hansteen} has a grid spacing of 100 km. In recent years, however, there has been a concerted effort to produce large-scale simulations with increasingly finer resolution. The enhanced network simulation by \citet{2016_Carlsson} has grid spacings of 48 km horizontally and 19 km vertically. More recently, a grid size of 23 km horizontally and 20 km vertically was achieved in \citet{2022_Przybylski,2024_Ondratschek}. The coronal bright point model by \citet{2023_Nobrega_Siverio} is resolved by 62.5 km in the horizontal direction and 50 km in the vertical direction. Similarly, the quiet Sun simulations described in \citet{2021_Chen} have a resolution of $\approx$ 48.8 km and 25 km in the horizontal and vertical directions, respectively. More idealized setups mimicking large-scale coronal loops exist with a minimum grid spacing of 30 km \citep{2023_Cozzo,2024_Cozzo}. Coronal loop simulations with a grid spacing as small as 12 km are described in \citet{2023_Breu,2024a_Breu}.
With new instruments such as the Extreme Ultraviolet Imager (EUI; \citet{2020_Rochus}) on Solar Orbiter \citep{2020A&A...642A...1M} or the future Multi-slit Solar Explorer (MUSE; \citet{2020_De_Pontieu, 2022_De_Pontieu})  the spatial resolution capabilities of observing instruments encroach upon or surpass the resolution of existing numerical simulations. At perihelion just inside 0.3 au, EUI can achieve a resolution of about 200 km \citep{2023_Berghmans}. The MUSE context imager aims at a resolution of about 240 km, a scale resolved by only a few grid points in many existing simulations. An increase in the resolution of observing instruments thus places strong demands on the numerical resolution of simulations.\\
Several existing MHD codes, e.g. MURaM \citep{2003_Vogler, 2004_Vogler, 2017_Rempel}, Bifrost \citep{2011_Gudiksen}, RAMENS \citep{2015_Iijima,2016_Iijima,2017_Iijima}, Mancha \citep{2006_Khomenko,2012_Khomenko,2010_Felipe,2018_Gonzalez, 2022_Navarro,2024_Modestov},  are now able to simulate the coupled system of convection zone, chromosphere, and corona in a single computational box.
Understanding how the convection zone, chromosphere and corona are coupled forms a substantial part of the coronal heating problem. All three regions are affected in different ways by the numerical resolution. The operation of the small-scale turbulent dynamo (SSD) in the convection zone and consequently the strength and distribution of the photospheric magnetic field is affected by the grid resolution \citep{2014ApJ...789..132R}. Estimates of the Poynting flux injected into the atmosphere have often focused on the motions of magnetic concentrations on the solar surface \citep{2015PASJ...67...18W}.
In addition to these bulk motions, small-scale motions within intergranular lanes could play an important role for coronal heating \citep{van_Ballegooijen_2011, 2020_Yadav, 2021A&A...645A...3Y, 2022_Breu, 2023_Breu, 2023_Kuniyoshi}. 
\citet{2020_Yadav} have shown that the Poynting flux might be significantly underestimated if the motions within the intergranular lanes are not sufficiently resolved. These structures, however, have widths on the order of only 100 km and are not properly resolved in many existing large-scale simulations. \\
All processes from energy injection to dissipation are affected by resolution. In his seminal paper on coronal heating via topological dissipation, \citet{1972ApJ...174..499P} suggested that infinitely thin current sheets will develop in a 3D magnetic field that is driven at the boundaries. The released energy depends on current sheet properties such as current sheet thickness. As infinitely thin current sheets can never be realized in a simulation that is based on a grid with a finite spatial resolution, numerical resolution influences energy release.\\
This results in a dependence of thermodynamic properties such as heating rate, temperature and density on the numerical resolution, with consequences for synthetic observables, since optically thin emission is very sensitive to the plasma density and the temperature distribution. Underestimating the plasma density therefore leads to a significant error in the synthesized emission.\\
Accurately capturing the exchange of material between the chromosphere and corona is vital to estimating coronal densities correctly.
Resolving the narrow chromosphere-corona transition region in large-scale 3D simulations poses a serious challenge. In the transition region, both temperature and density exhibit steep gradients over a short length scale, requiring very small grid spacing \citep{2015_Peter}. If the transition region is underresolved, the coronal density will be underestimated by a factor of up to two. This effect has been studied in e.g. \citet{Bradshaw_2013}. The resolution of the transition region also plays an important role in determining how much energy is injected into the corona, depending on how well perturbations propagating through it are resolved \citep{2023MNRAS.526..499H}.\\
It should be noted that the coronal density in numerical simulations does not only depend on numerical resolution, but also on the inclusion of other effects, such as non-equilibrium ionization determining the density of ions responsible for coronal emission. \citet{1993_Hansteen} found that the optically thin radiative loss curve can vary by a factor of more than two over time due to plasma flows and waves if out-of-equilibrium effects are included. The effect of non-equilibrium ionization on EUV emissivity was further studied in \citet{2003_Bradshaw} and \citet{2011_Bradshaw}. This has consequences especially for the detection of emission from very hot plasma due to short heating timescales predicted by the nanoflare model of heating. The temperature changes are then too fast for the ionization state to collisionally adjust.\\
The inclusion of non-equilibrium effects is especially important to reproduce observed UV emission lines originating from the transition region, as pointed out in \citet{2016_Martinez_Sykora}, \citet{2018_Nobrega_Siverio} and \citet{2019_Bradshaw}.
In our study, however, we focus on the coronal EUV emission and the corona is assumed to be in ionization equilibrium and the inclusion of non-equilibrium effects, while likely having an important effect on especially very hot EUV emission, is out of the scope of this work.
\\
Emission of coronal loops is observed to be structured in threads on scales of a few hundred to about 2000 km \citep{2017_Aschwanden}. 
Heat conduction in magnetized plasmas is highly anisotropic, operating far more efficiently along the magnetic field than in the cross-field direction. Variations in heating transverse to the magnetic field will therefore lead to elongated bright structures aligned with the magnetic field. Whether existing instruments are capable of resolving the substructure of these coronal loop strands is an open question \citep{2013_Peter, 2020a_Williams, 2020b_Williams}. In a numerical simulation, the spatial resolution will also affect the smallest resolvable substructures.\\
Since plasma flows on a range of scales in the atmosphere shift and broaden emission line profiles, the shape of the line profile is influenced by spatial resolution in addition to the intensity.
Resolving small-scale velocity fluctuations is important to model emission line widths accurately \citep{2024a_Breu}.\\
Given how many aspects of coronal simulations are significantly influenced by resolution, it is vital to quantify the effect on predictions for observables derived from numerical simulations in order to decide how much trust to place on forward modelling results.
Since numerical simulations in high resolution are costly in terms of computational time, accuracy has to be weighed against information gain about physical processes and resource use. The aim of this paper is to quantify how much and what information we are losing from simulations when conducting them with a lower grid spacing. With our choice of grid spacings for out numerical models we bridge the gap from the (large-scale) active-region models to high-resolution models resolving the small-scale magnetic evolution in the photosphere, in particular in intergranular lanes. Hence, the latter models should at least be reliable in terms of the energy supply to the corona.\\
In this paper, we examine the effects of resolution on thermodynamic quantities in the coronal loops, derived synthetic observables, and observable structure sizes in the corona.
We will introduce the numerical setup and analysis methods in section \ref{sec:methods}, describe the results in section \ref{sec:results}, followed by a discussion in section \ref{sec:discussion} and conclusions in section \ref{sec:conclusions}.

\section{Methods}
\label{sec:methods}

\subsection{Simulation Setup}
\label{sec:setup} 

\begin{figure*}
 \resizebox{\hsize}{!}{\includegraphics{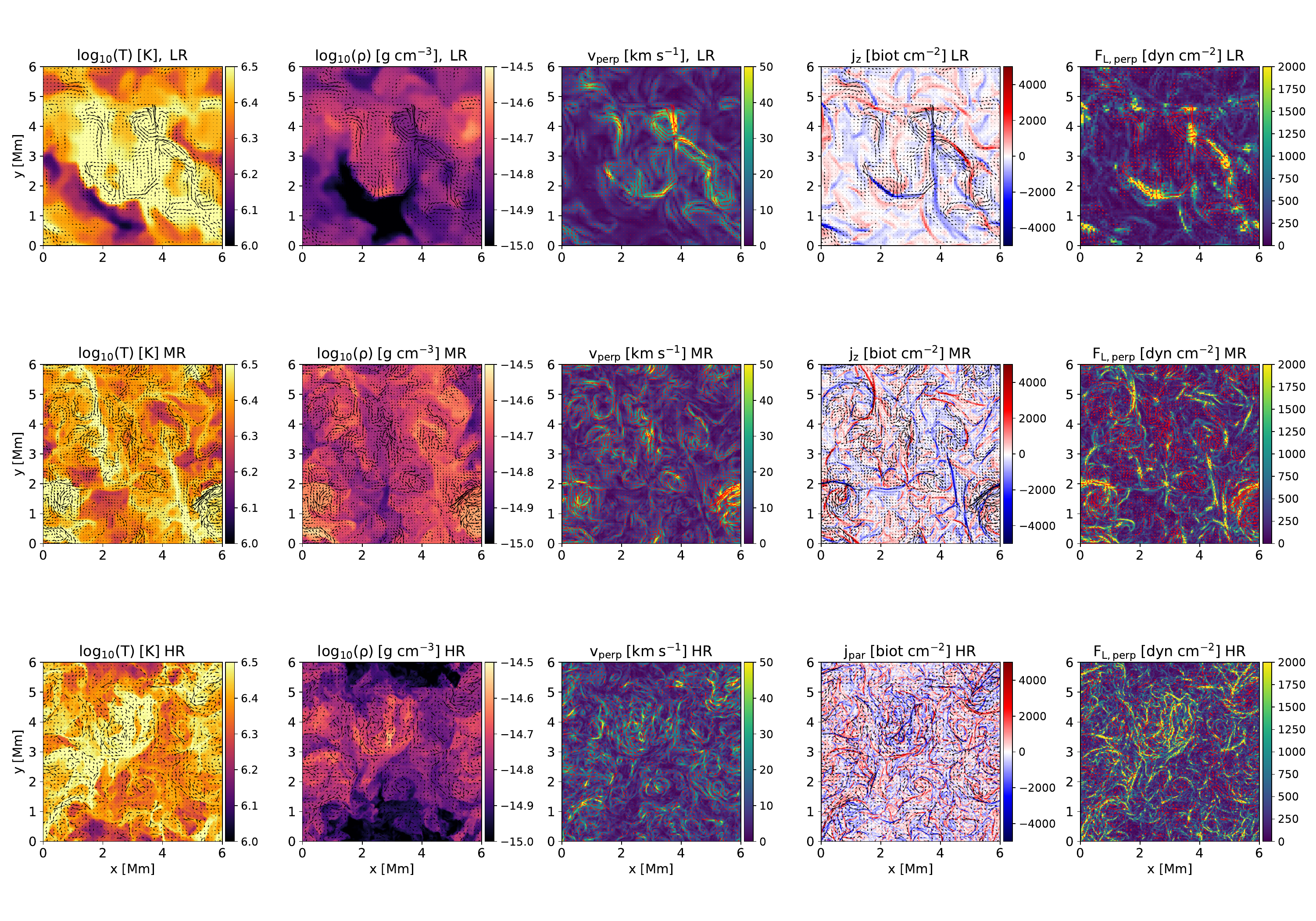}}
    \caption{Cut through the simulation box at the apex for three different resolutions. The simulation time instance corresponds to 50.2 min for the LR simulation, 40.1 min for the MR simulation, and 17.6 min for the HR simulation.} From top to bottom: Low, medium and high resolution. From left to right: Temperature, mass density, velocity perpendicular to the loop axis, current density parallel to the loop axis, magnitude of the Lorentz force perpendicular to the loop axis. The black arrows illustrate the in-plane velocity field.
    \label{fig:ap_cut_difres}
\end{figure*}

We perform 3D resistive MHD simulations of a coronal loop for different numerical resolutions with the 3D radiative MHD code MURaM. MURaM solves the compressible MHD equations for a partially ionized plasma using a fourth-order finite-difference scheme in addition a short characteristics scheme for the radiative transfer using\citep{2003PhDT........61V, 2005A&A...429..335V}.\\
The coronal loop is modelled as a straightened-out magnetic flux tube including a shallow convection zone layer at each footpoint and a coronal layer in between. 
The coronal extension of the MURaM code is described in detail in \citet{2017_Rempel}, while the stretched loop setup is introduced in \citet{2022_Breu}.
The simulation domain includes a shallow convection zone, a chromosphere in local thermodynamic equilibrium and a coronal layer between both footpoints. Effects from gravitational stratification, field-aligned Spitzer heat conduction, optically thick radiative transfer in the photosphere and chromosphere and optically thin radiative losses in the corona are included.
The initial magnetic field is chosen as a vertical uniform field with a strength of 60 G. 
Three different simulation runs are conducted with resolutions $\Delta_{LR}=60\; \rm{km}$, $\Delta_{MR}=24\; \rm{km}$ and $\Delta_{HR}=12\; \rm{km}$. In the following, we will refer to the runs with different resolutions as LR, MR and HR for Low, Medium and High Resolution. The initial condition for the medium and high-resolution runs was interpolated from a snapshot of the low-resolution run. The simulation box was then evolved for another 30 minutes of solar time to let the initial transients subside before data was taken. \\
The simulation box is covered by a grid with $100 \times 100 \times 950$ points for the LR simulation,
$250 \times 250 \times 2375$ points for the MR run and $500 \times 500 \times 4750$ points for the HR run. The side boundaries are periodic. The dimensions of the simulation box are $6\times 6\times 57$ Mm, with a convection zone depth of 3.5 Mm leading to an effective loop length of 50 Mm. A cut through the loop apex for the three different resolutions is shown in Fig. \ref{fig:ap_cut_difres}.\\
Due to the high computational cost and required storage space for snapshots, available time series for the MR and HR runs are shorter (40 min and 70 min vs 85 min) and were written out with a lower temporal cadence than the LR run.  While the cadence of the LR run is 28 seconds, the snapshots of the MR and HR run are written out with a cadence of 102 and 66 seconds, respectively.

\subsection{Synthesized Emission}

The optically thin emission is given by $\epsilon=n_{e}^{2}K(T)$, where $n_{e}$ is the electron density and $K(T)$ is a response function depending on the temperature. The kernel $K(T)$ takes into account the contribution function of the optically thin emission lines and the effective area of the observing instrument. Here we use the response function of the Atmospheric Imaging Assembly (AIA) \citep{2012SoPh..275...17L} and of the XRT instrument on board of the Hinode spacecraft \citep{2007_Golub}. We neglect the slight density dependence of the emission line response function and treat it purely as a function of temperature. The emission is then first computed from the density and temperature at every grid point. In order to obtain the synthetic observations, we then integrate along the line-of-sight, here chosen as the direction perpendicular to the loop axis. 
The dependence of the response function on the electron density originates in shifts of the ionization equilibrium with density, i.e. the peak formation temperature of a specific ion will change slightly if the electron density changes. However, the exact form of the density dependence varies for different ionisation models \citep{2020_Dufresne}.
While the AIA response function does have a slight dependence on the electron density, neglecting this dependence was found to lead to errors of at most 10 \% for most pixels in forward modelling and mostly affects small-scale transition region structures (Yajie Chen, private communication).

\subsection{Coronal strand widths}

\begin{figure*}
    \resizebox{\hsize}{!}{\includegraphics{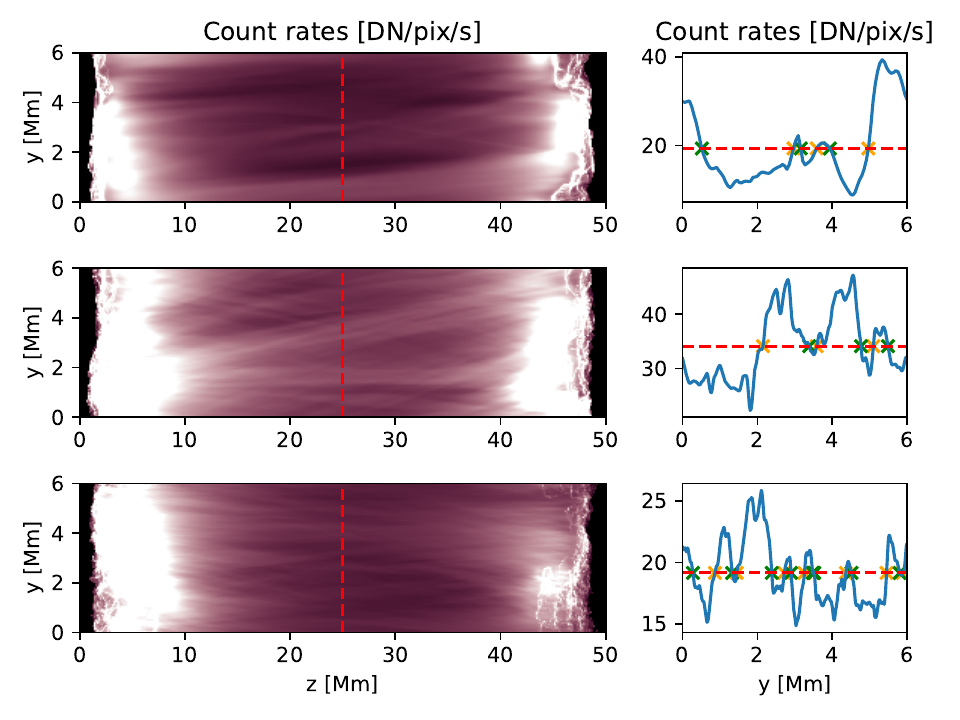}}
    \caption{Left: Integrated emission in the 211 \AA\; channel of AIA. Right: Cut through the emission at the apex. The red dashed line is the threshold used for the strand width computation. The orange and green dots mark the start and endpoint for the computation of the width of each peak.}
    \label{fig:AIA211}
\end{figure*}

While observed coronal loops appear to be structured in the direction transverse to the loop axis, the concept of a coronal loop strand is not very clearly defined, and there is no unique method for the identification of the strand widths. Since coronal emission is optically thin, observed coronal structures suffer from projection effects. The question whether compact bright structures in the corona exist or if the observed structuring arises from projection effects of highly corrugated emitting plasma structures is the subject of ongoing debate \citep{2022_Malanushenko, 2024_Uritsky, 2024_Mandal, 2024_Ram}.\\
In order to quantify the scale of structures in the optically thin emission from the loop, we employ two different methods. The first method is to compute a power spectrum of the emission in different filters at the apex. The power spectrum is computed as
\begin{equation}
    E(k)=\frac{1}{N_{t}}\sum_{j=0}^{N_{t}}\int_{k\leq k+\Delta k} \vert \widehat{\sqrt{I}}\vert^{2}dk,
\end{equation}
where $\widehat{\sqrt{I}}$ is the Fourier transform of the square root of the intensity and $N_{t}$ is the number of time steps the Fourier spectrum is averaged over, while $k$ is the spatial wavenumber.\\

The second method we use to determine the size of structures in the coronal emission follows \citet{2024_Uritsky}. Here, a constant detection threshold equal to the mean value of the count rate at the loop apex is applied and the intersection of the peaks in emission with the threshold level is determined. The strand width is then computed as the distance between intersections. This method is illustrated in Fig. \ref{fig:AIA211}.

\section{Results}
\label{sec:results}

In the following subsections, we analyze power spectra as well as time series of averaged coronal quantities. We also investigate synthesized observables and the 3D structure of different atmospheric layers.
First, we discuss the effect on energy injection, thermodynamic bulk properties and synthesized emission in section \ref{sec:thermo}. Then we quantify the effect of resolution on the stratification of different atmospheric layers in section \ref{sect:tr_dyn}.

\subsection{Thermodynamic Quantities and Emission}
\label{sec:thermo}

\begin{figure*}
\resizebox{\hsize}{!}{\includegraphics[width=\columnwidth]{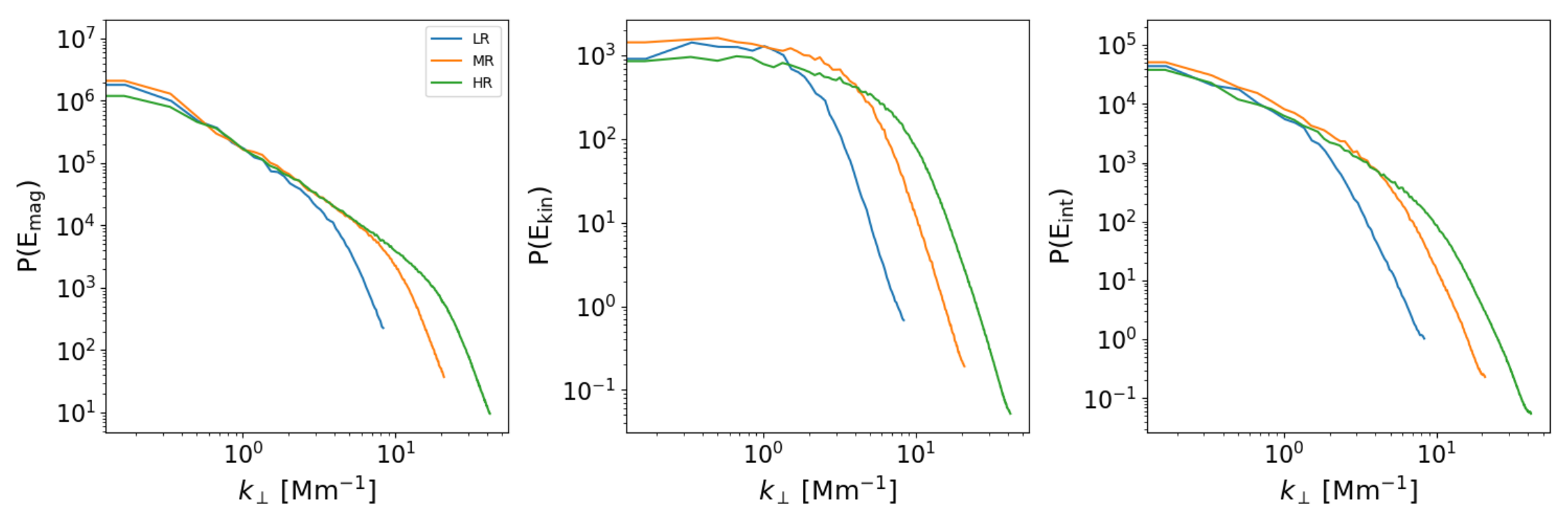}}
    \caption{Spatial power spectra for the in-plane magnetic, kinetic and internal energy densities in a cut at the loop apex for three different resolutions.}
    \label{fig:psd_energies}
\end{figure*}

\begin{figure*}
\resizebox{\hsize}{!}{\includegraphics{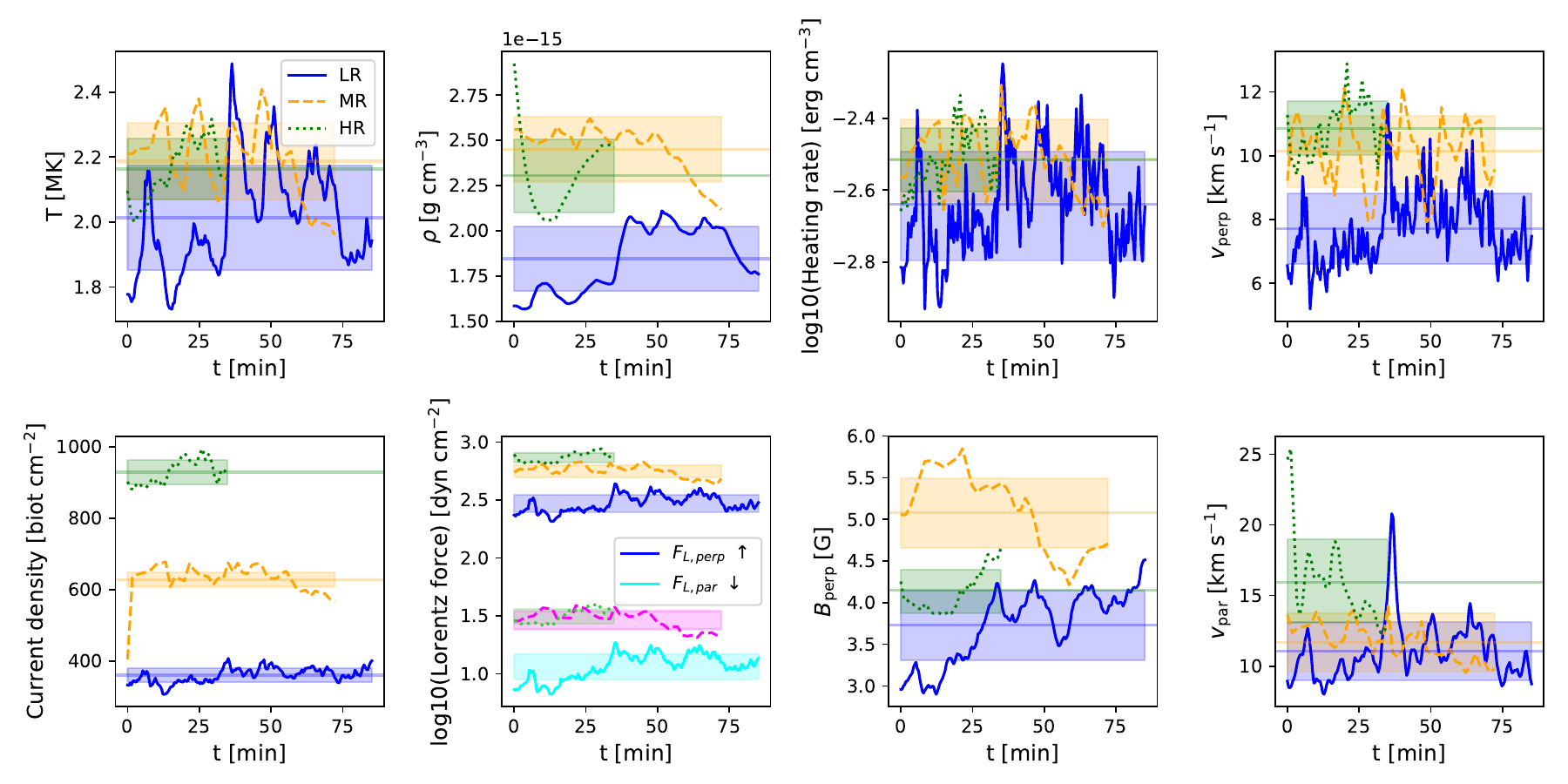}}
    \caption{Time evolution of a selection of thermodynamic quantities averaged in regions with temperatures above $10^5$ K for different resolutions.  Top row: Average coronal temperature, mass density, heating rate and the velocity component perpendicular to the loop axis. The heating rate is computed as the sum of resistive and viscous heating. Bottom row: average current density, average Lorentz force perpendicular (blue, orange, green) and parallel (cyan, magenta, light green) to the loop axis, magnetic field component perpendicular to the loop axis and absolute value of the velocity component parallel to the loop axis. Solid horizontal lines show the value of the respective quantities averaged over the whole time series, while the shaded areas illustrate the standard deviation obtained from the temporal variations of the respective quantities.}
    \label{fig:quant_evo}
\end{figure*}

\begin{figure*}
\resizebox{\hsize}{!}{\includegraphics{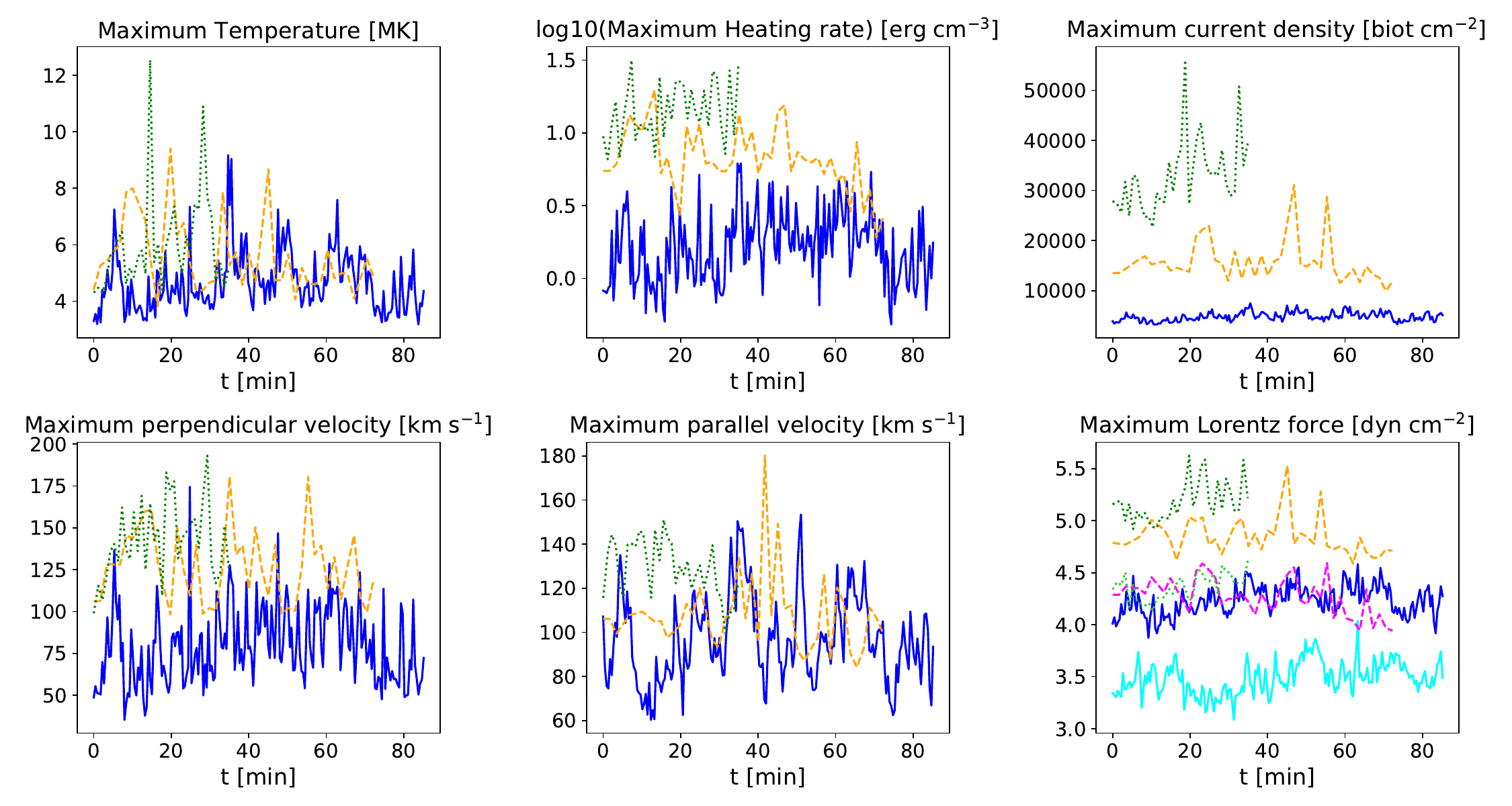}}
    \caption{Time evolution of the maximum value of several thermodynamic quantities in regions with temperatures above 100 000 K for different resolutions.  Top row:  Coronal peak temperature, maximum total heating rate and maximum current density. Bottom row: Maximum velocity perpendicular to the loop axis, maximum magnitude of the velocity parallel to the loop axis and maximum of the Lorentz force perpendicular (blue, orange, green) and parallel to the loop axis (cyan, magenta, light green).}
    \label{fig:quant_max_evo}
\end{figure*}

\begin{figure}
\includegraphics[width=\columnwidth]{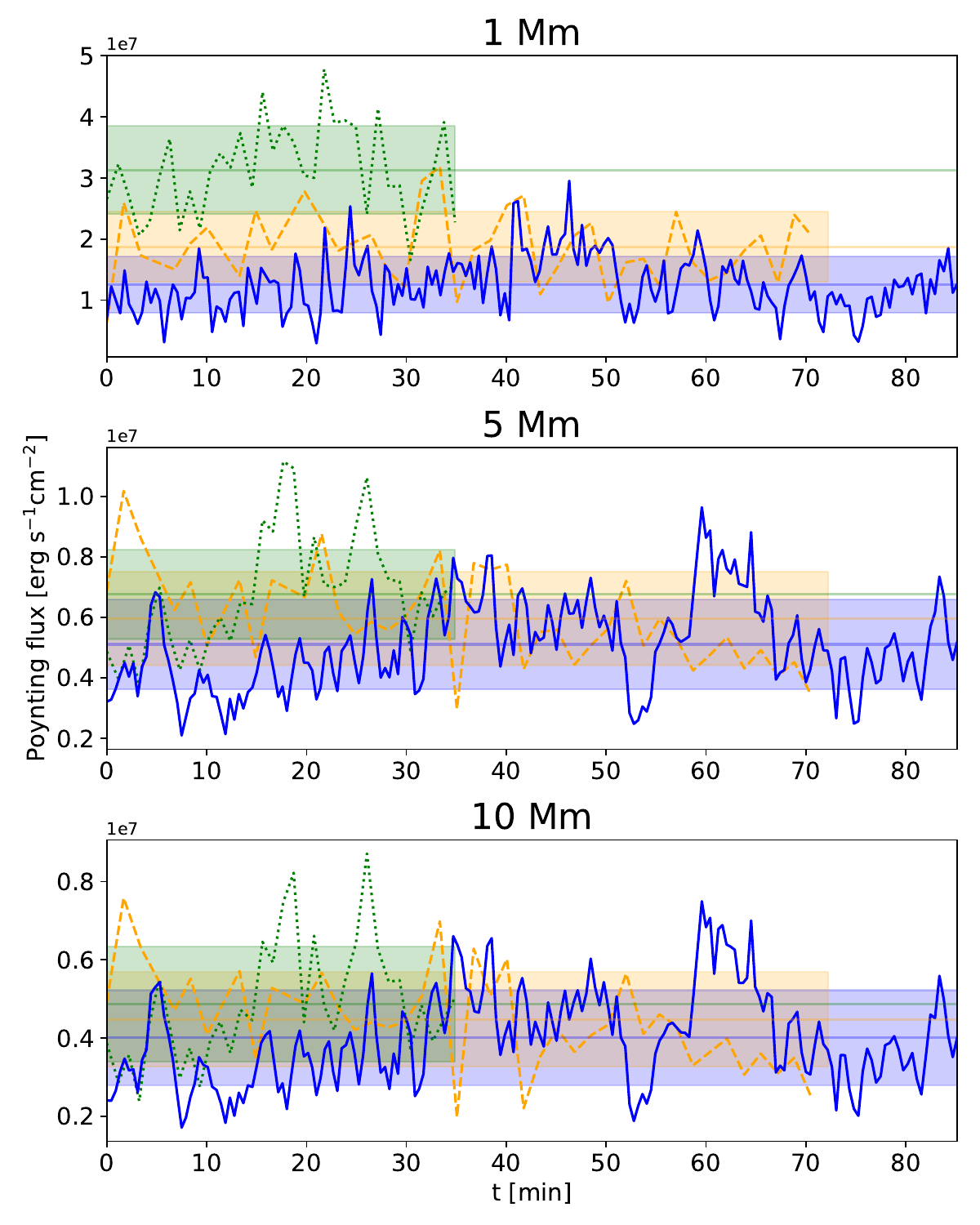}
    \caption{Time series of the average Poynting flux through a slice at 1 Mm above the photosphere (top row), 5 Mm (middle row) and 10 Mm (bottom row) for different numerical resolutions (LR: solid, MR: dashed, HR: dotted). The Poynting flux was averaged over both footpoints, taking into account the opposite sign of the injected Poynting flux. The shaded areas correspond to the standard deviation of the temporally averaged Poynting flux.}
    \label{fig:poynt_difres}
\end{figure}

\begin{figure}
\includegraphics[width=\columnwidth]{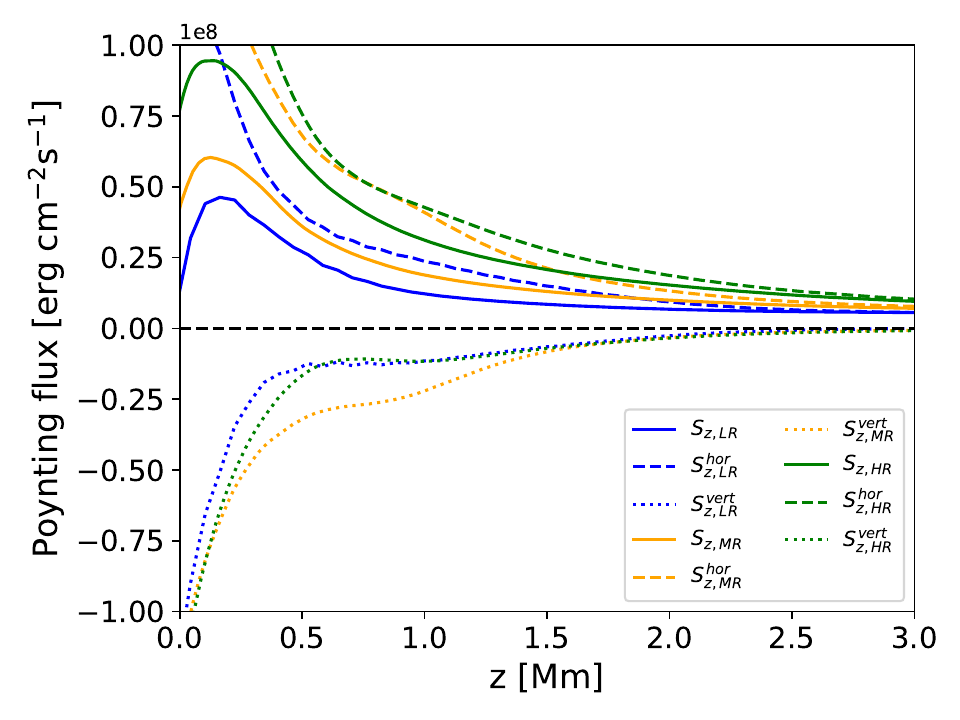}
    \caption{Chromospheric Poynting flux averaged over the loop cross section for the LR, MR and HR run. The solid lines show the total Poynting flux, while the dashed lines denote the component due to the action of horizontal flows on the magnetic field and the dotted lines the vertical advection of horizontal magnetic field.
    The Poynting flux was averaged over all snapshots from the analysed  time range for each run and over both footpoints, taking into account the opposite sign at each footpoint.}
    \label{fig:poynt_chromo}
\end{figure}

\begin{figure}
\includegraphics[width=\columnwidth]{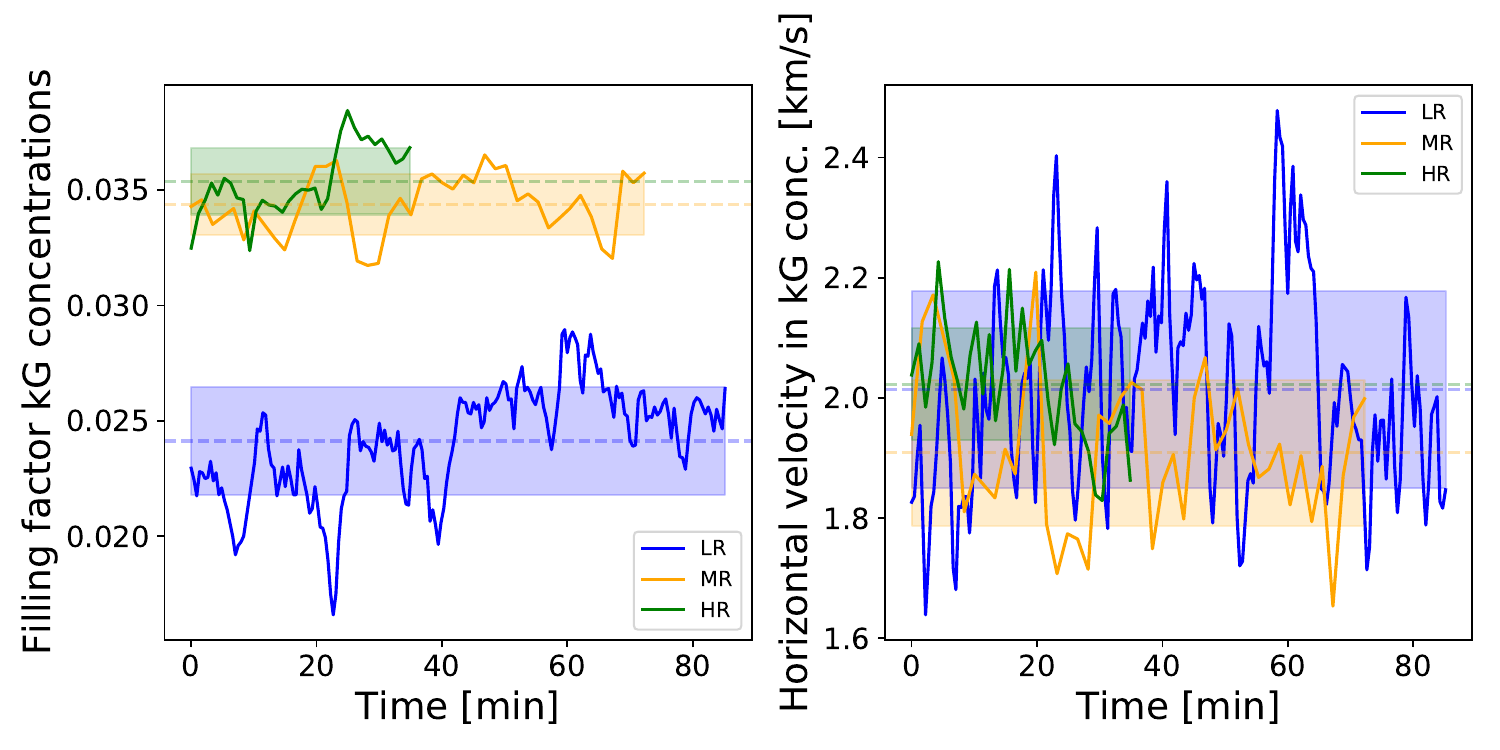}
    \caption{Left panel: Filling factor of kilogauss magnetic field concentrations in the photosphere for different resolutions. Right panel: Average horizontal velocity inside kilogauss magnetic field concentrations. The horizontal dashed lines mark the respective time-averaged values for each run. The shaded areas correspond to the standard deviations of the temporally averaged quantities.}
    \label{fig:fill_fact}
\end{figure}

\begin{figure}
\includegraphics[width=\columnwidth]{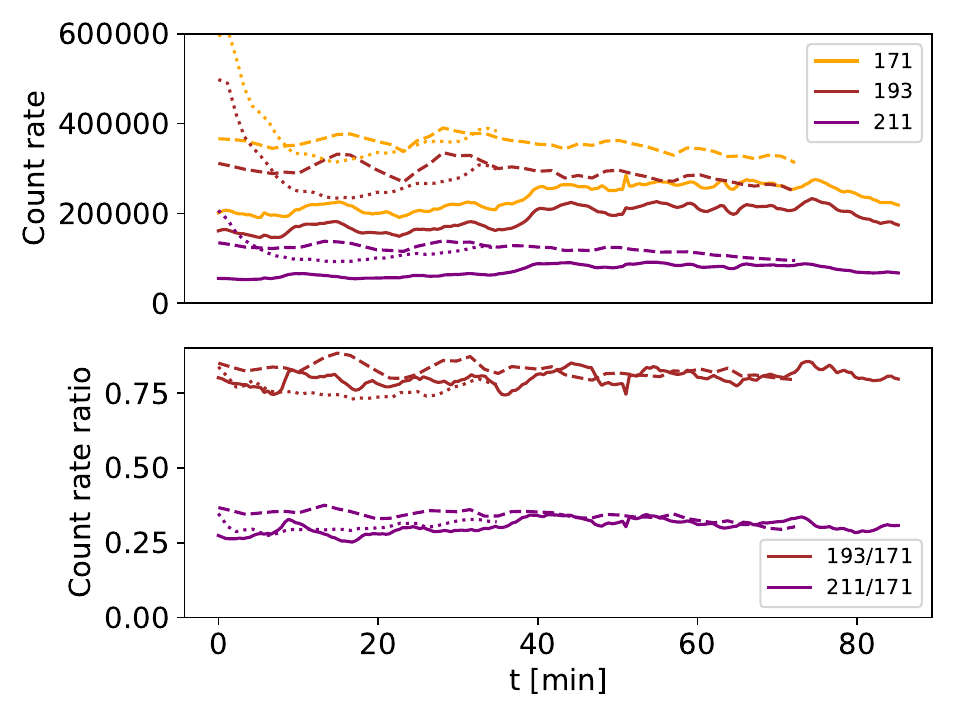}
    \caption{Time series of emission detected with AIA for different wavelength channels and numerical resolutions. Solid line: LR, dashed line: MR, dotted line: HR. Top panel: Time evolution of count rates for different channels. Bottom panel: Time evolution of count rates for the 193 and 211 \AA\ channel normalized by the total emission detected in the AIA 171 \AA\ channel.}
    \label{fig:aia_ts}
\end{figure}

Coronal emission depends on temperatures and densities in the corona and therefore on the energy input. Fig. \ref{fig:ap_cut_difres} demonstrates the influence of numerical resolution on various quantities in a plane at the loop apex. While the distribution of temperature and density shows finer spatial structures, the effect of spatial resolution is most noticeable for the velocity field and the current distribution. The velocity field is displaying increasingly smaller and more space-filling eddies. While the low-resolution simulation contains a few strong current sheets, the current sheets become smaller, but stronger and more numerous with increasing resolution. For the HR run, almost the entire loop cross section is filled with small-scale current sheets. The behaviour of the current sheet distribution is mirrored in the Lorentz force distribution in the coronal volume, showing increasingly smaller spatial scales for the MR and HR run. Stronger current sheets lead to stronger Lorentz forces and thus also to stronger small-scale flows in the loop.\\ 

\subsubsection{Spatial power spectra of energies in the loop}

We compare the influence of the resolution on the energy content of the loop and the spatial distribution of different types of energy.
Energetically, the corona is dominated by the magnetic field, implying that the kinetic energy in the corona is much smaller than the magnetic energy. With increasing resolution, we resolve smaller structures in the magnetic field and flow field. While we expect the energy power spectra to be cut off at smaller spatial wavenumbers for lower resolutions, it is important to test whether the spectral distributions of magnetic, kinetic and internal energy behave similarly at large scales and only differ regarding the cutoff at small scales, or if the choice of numerical resolution also affects larger scales by redistributing energy from small to large scales.\\
Time-averaged spatial power spectra for the in-plane magnetic energy, kinetic energy, and internal energy density at the loop apex are plotted in Fig. \ref{fig:psd_energies}. The spectra are averaged over 10 snapshots each. We find that the magnetic energy spectrum shows clear power-law behaviour for the MR and HR runs, which is less pronounced for the lowest resolution case. For small wavenumbers, there is slightly more energy present for the LR and MR runs than for the HR run for the magnetic and kinetic energy densities. The internal energy density does not show a significant difference at small wavenumbers. The numerical resolution does therefore not only influence the distribution of magnetic and kinetic energies at high wavenumbers, but also the behaviour of the magnetic field and especially the kinetic energy at larger spatial scales. The magnetic, kinetic and internal energies in the loop midplane do not increase monotonically with decreasing grid spacing, although the MR and HR runs have higher average energy densities in the loop midplane than the LR run for magnetic, kinetic and internal energy densities.\\ 
Magnetic and kinetic energies fluctuate quite strongly over the course of the time series and runs of durations of several hours for each grid size would be necessary to properly test for convergence. This is not feasible for the HR run due to the high computational cost. The MR run is in a state of high magnetic energy due to two sheared flux bundles developing over the course of the run. If the HR simulation was run for longer, it might reach an even higher magnetic energy.

\subsubsection{Time evolution of average quantities}

Resolution not only affects the spatial distribution of energies, but also average properties of the loop. In the following, we examine the time evolution of average coronal quantities. %
Here we define the coronal volume as regions with temperatures above $10^5\,$K. Plasma hotter than $10^5\,$K takes up a comparable fraction of the atmosphere above the photosphere. On average, the fraction of coronal plasma is 91.6 \%, 91.4 \% and 89.9 \% for the LR, MR and HR run, respectively. The coronal fraction of the simulation box does not change significantly over the course of the examined time series.
The results are summarized in Fig. \ref{fig:quant_evo} and Fig. \ref{fig:quant_max_evo}. We find that both mean and peak temperature are resolution dependent, with a slight increase of the average coronal temperature with resolution. While the time-averaged coronal temperature is more than $10^{5}$ K higher for the MR and HR runs than for the LR run, the difference between the MR and HR runs is small. The time-averaged temperature is slightly higher for the MR run. The highest temperature value measured over the entire time series, however, does increase monotonically with numerical resolution.
While even for the LR run local temperatures of up to 9.2 MK can be reached, we find the peak temperature for the MR run to be 9.4 MK and 12 MK for the HR run. Thus, even in the low-resolution run, the plasma locally reaches flare temperatures. The relatively small influence of the resolution on the average coronal temperature, however, indicates that the high-temperature plasma has a very small filling factor.\\
Similar to the temperature, the density increases with resolution from LR to MR. Due to the limited amount of data available for the medium and especially for the high resolution run and the strong fluctuations of the average coronal density in time, it cannot be determined from the available timeseries whether there is a further increase in average density for the HR run compared to the MR run. The HR loop is in a draining phase in the time range over which data was taken and on average has lower densities than the MR run.

\subsubsection{Heating rates and Lorentz force}

Resistive and viscous heating rate as well as current density increase with resolution, leading to increased temperatures. The spatially and temporally averaged heating rate is almost the same between the MR and HR runs. The HR run, however, reaches the highest maximum values in the coronal volume for both the heating rate and the current density.\\
Consistent with Fig. \ref{fig:psd_energies}, the velocity perpendicular to the loop axis is generally higher for the MR and HR runs, with the highest average and peak values reached for the HR run. Due to the low plasma beta in the corona, the plasma is confined to moving along magnetic field lines. Therefore we expect the Lorentz force to be the main factor responsible for accelerating the plasma in the direction perpendicular to the loop axis.The Lorentz force depends both on the current density and the magnetic field strength. While the magnetic field component perpendicular to the loop axis is on average higher for the HR run than for the LR run, it fluctuates strongly in time and can reach comparable values for both simulations. In the stretched-loop setup, the transverse component of the magnetic field can be increased either due to a higher degree of small-scale tangling and twisting of the magnetic field or due to large-scale inclination of the magnetic field in the simulation box. The MR run shows higher average values for the perpendicular magnetic field than the HR run, therefore we do not find a systematic behaviour of the perpendicular magnetic field with grid resolution. In contrast, the maximum current density is significantly higher for the HR run. Higher maximum values for the current density will lead to a stronger Lorentz force. We find a higher average and maximum component of the Lorentz force directed perpendicular to the loop axis for the MR and HR runs, with the highest values for the HR run. Consequently, the plasma experiences a stronger acceleration near current sheets. The component of the Lorentz force parallel to the loop axis is also higher for the MR and HR runs, although it does not vary significantly between the MR and HR simulations.

\subsubsection{Poynting Fluxes}

Consistent with the higher energy content of the MR and HR runs, we find that a larger Poynting flux is injected into the loop for higher grid resolutions. The Poynting flux as a function of time for different slices along the loop is shown in Fig.~\ref{fig:poynt_difres}. The difference between runs with different grid resolutions is largest in the chromosphere and gets smaller with increasing height above the photosphere. The increased Poynting flux could either originate from better resolution of small-scale magnetic field structures and intergranular motions or a higher horizontal magnetic field component at the coronal base. The difference in Poynting flux between the simulation runs is most pronounced in the chromospheric layer (see the upper panel of Fig. \ref{fig:poynt_difres}), indicating that more Poynting flux is available at the solar surface and injected into the atmosphere instead of a larger fraction of Poynting flux reaching the corona.\\
The total vertical Poynting flux can be decomposed as follows \citep{2012_Shelyag}:
\begin{equation}
    S_{z} = \frac{1}{4\pi}(v_{z}(B_{x}^{2}+B_{y}^{2}) - B_{z}(v_{x}B_{x} + v_{y}B_{y})).
\end{equation}
Here the first term is the component produced by the transport of horizontal magnetic field by vertical plasma flows (flux emergence) and the second term is generated by horizontal plasma flows acting on the inclined magnetic field. The two components of the Poynting flux and the total Poynting flux in the chromosphere are shown in Fig. \ref{fig:poynt_chromo}. The Poynting flux generated by horizontal plasma motions starts to dominate above the photosphere. While the term caused by the vertical transport of horizontal magnetic field shows no systematic behaviour, the Poynting flux generated by horizontal motions increases with resolution for the grid sizes tested, leading to the same behaviour in the chromospheric total Poynting flux.
The average unsigned magnetic field at the photosphere is 82.6 G for the LR run, 106 G for the MR run and 108 G for the HR run, despite all runs starting with an initial uniform magnetic field of 60 G. The simulations with smaller grid spacing resolve more of the small opposite-polarity magnetic field patches generated by magnetoconvection, leading to a larger unsigned magnetic field strength.
The filling factor of magnetic concentrations with kilogauss field strength in the photosphere increases with resolution as shown in Fig. \ref{fig:fill_fact}, although the difference between the MR and HR runs is small. The vertical Poynting flux at the photosphere depends on both magnetic field strength and velocities. We computed the horizontal velocities $\langle v_{\rm{hor}}\rangle_{kG}$ averaged over grid cells with a vertical magnetic field strength above 1000 G at a height where $\langle \tau\rangle$=1 for each snapshot. For all resolutions, the time-averaged velocities have a magnitude of roughly $2\; \rm{km\; s^{-1}}$. This is slightly higher than the measured rms velocities of magnetic bright points of 1.34 $\rm{km\; s^{-1}}$ \citep{2012ApJ...752...48C}. This suggests that the Poynting flux increase with resolution is due to resolved small-scale magnetic field structures in the MR and HR runs, not larger velocities within those concentrations.\\
Both energy flux and mass transport are affected by resolution.
The vertical velocity reaches comparable maximum values in the coronal domain for all three resolutions, although the average  vertical velocity is larger for the HR run. Calculating the mass flux through a slice at a constant height, we find that the mass flux reaches larger absolute values for the low-resolution simulation. There is, however, a higher number of peaks in a comparable time interval in the MR and HR time series, suggesting a more dynamic atmosphere with more frequent up- and downflows of dense material. We will investigate this behaviour further in sect. \ref{sect:tr_dyn}.\\

\subsubsection{EUV emission}

Finally, we study the impact of the elevated coronal energy content with increased numerical resolution on magnitude and structure of forward-modelled emission from the models.  We compute synthetic observables as seen by AIA in different channels for all three runs.
The optically thin emission from the corona is both temperature and density dependent, therefore we expect synthetic observables to also depend on numerical resolution. The average count rates as a function of time are shown in Fig.~\ref{fig:aia_ts}. We find that the curves follow the evolution of the average coronal density.
As expected from the behaviour of temperature and density, the average count rate increases with resolution.
While there is a significant jump in measured count rates between the low and medium resolution runs, the difference is less pronounced for medium and high resolution.
Since the temperature also increases slightly with resolution, one might expect that the emission is shifted to hotter channels and a larger fraction of the plasma will show up in hotter AIA channels. The emission in the 193 \AA\; and 211 \AA\; channels normalised by the emission in the 171 \AA\; channel for each time step is shown in the bottom panel of Fig. \ref{fig:aia_ts}.
We find that the emission in all AIA channels increases, while the relative emission between different channels does not strongly depend on resolution.\\
\begin{figure*}
\resizebox{\hsize}{!}{\includegraphics{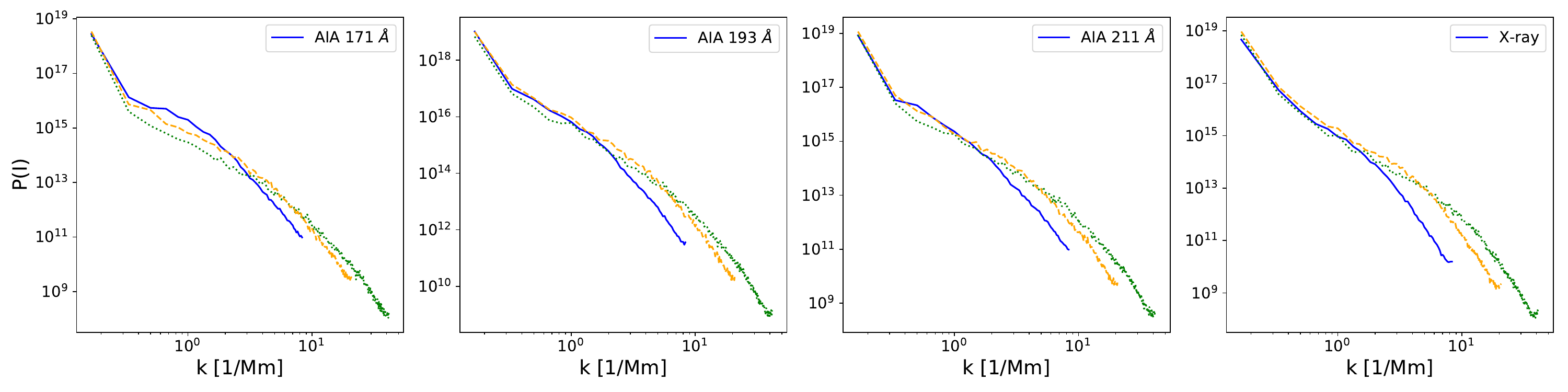}}
\caption{Power spectra for AIA and X-ray emission at the loop apex for different grid resolutions (LR: solid, MR: dashed, HR: dotted). From left to right: 171 \AA\, 193 \AA\ and 211 \AA\ wavelength channels for AIA, Al-poly filter for Hinode XRT.}
    \label{fig:psd_strands_AIA}
\end{figure*}

\begin{figure*}
\resizebox{\hsize}{!}{\includegraphics{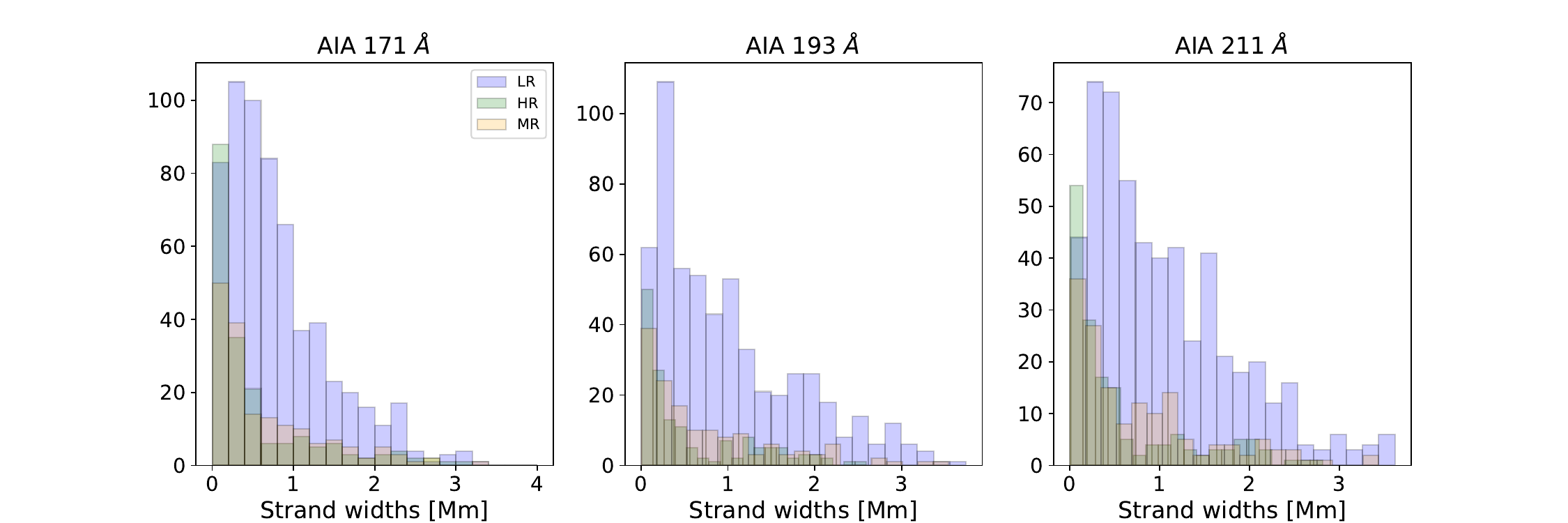}}
    \caption{Histogram of strand sizes for different AIA channels and numerical resolutions. From left to right: 171, 193 and 211 \AA\ channel for AIA.}
    \label{fig:histo_strands_AIA}
\end{figure*}

\subsubsection{Loop strand widths}

Given the ongoing debate over what determines the scale of a coronal loop strand \citep{2024_Uritsky} or if well-defined strands even exist \citep{2022ApJ...927....1M}, it is interesting to check whether the numerical resolution affects detected strand sizes or if there exists a "fundamental" coronal loop strand width. Fig. \ref{fig:AIA211} illustrates that while all loops show features in the emission on a range of scales, the MR and HR runs display more fine structure.
Computing a power spectrum of the emission in different AIA bands and the X-ray emission for a slit at the loop apex, we find that structures exist on all scales present in the simulations and a clear peak at a specific scale is absent as illustrated in Fig. \ref{fig:psd_strands_AIA}. The power spectral density is larger for small scales in the MR and HR runs, but the power spectra are similar for larger scales up to about $1\; \rm{Mm^{-1}}$, especially for the MR and HR runs. For the emission in different AIA channels, more power seems to be present at large scales for the low-resolution simulation. This is especially pronounced for the lowest temperature channel.\\
In addition to power spectra, we used a second method to determine coronal strand sizes by measuring their width at a constant detection threshold.
The resulting distribution of strand sizes is shown in Fig.\ref{fig:histo_strands_AIA}. While all runs produce structures on the scale of several Mm, the MR and HR runs show a larger number of fine-scale structures. The histograms for the LR run peaks around 0.5 Mm for all three AIA channels. In contrast to this, the histograms for the MR and HR runs peak for the lowest strand widths. The widest structures have widths around 3.5 Mm for all resolutions. The clear peak of the histograms for small strand sizes vanishes if the emission is degraded to the resolution of AIA by convolving with a 2D Gaussian, showing that the high-resolution simulation runs produce structure sizes below the resolving capabilities of current instruments.

\subsection{Dynamics of jets at the loop base}
\label{sect:tr_dyn}

\begin{figure}
\includegraphics[width=\columnwidth]{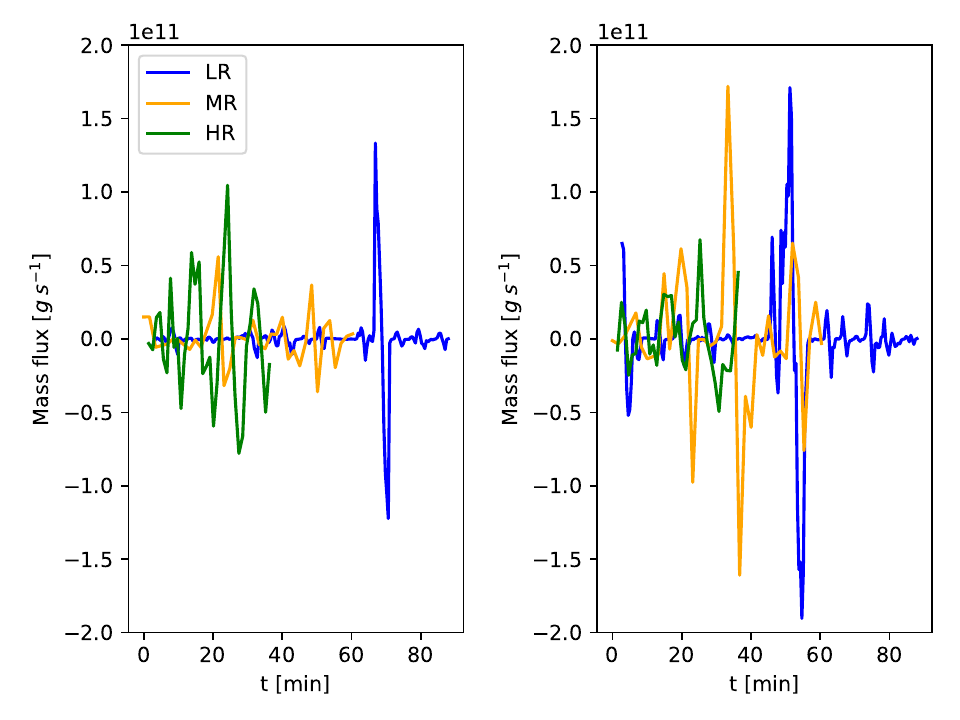}
    \caption{Time evolution of the net mass flux through a slice located at three Mm above the photosphere for footpoint 1 (left panel) and footpoint 2 (right panel) for different resolutions. }
    \label{fig:vel_mflux}
\end{figure}

While early models of the solar atmosphere assumed a plane-parallel stratification, observations \citep{2010NewAR..54...13T} and 3D MHD models have shown that the transition region is, in fact, highly corrugated \citep[e.g.][]{2005ApJ...618.1020G,2006ApJ...638.1086P,2017ApJ...848...38I}. While very thin in the vertical direction, it is expected that the transition region will appear much thicker if the emission is integrated along a horizontal LOS due to its convoluted nature \citep{2006ApJ...638.1086P}.\\
The three simulation runs studied in this paper differ quantitatively in the structuring of the lower atmosphere. With higher numerical resolution, the cooler chromospheric plasma reaches larger geometrical heights from the photosphere.\\
We define the transition region height as the height, measured from the photospheric level, at which the plasma reaches a temperature equal to $10^{5}$ K. We record the average and the maximum height of the transition region for each timestep. We find that the reason for the frequent excursions in the mass flux for the MR and HR simulations shown in Fig. \ref{fig:vel_mflux} is the presence of a larger number of jets of cool and dense chromospheric material crossing the slice perpendicular to the loop axis located at 5 Mm above the photosphere at which the mass flux was recorded for the MR and HR runs. Both the average and the maximum height of the transition region reached during the simulation run increase with numerical resolution.   While the highest jets for the LR run reach heights of about 5 Mm, the highest jet for the HR run is more than 7.73 Mm high and the highest jet in the MR run reaches 6.73 Mm. \\
\begin{table}
\caption{Mean, standard deviation and skewness for the height distribution of the transition region for the LR, MR and HR run.}
    \begin{center}
        \begin{tabular}{ c c c c}
              & LR & MR & HR\\ 
             \hline
             Mean [Mm] & 2.06 & 2.15 & 2.51\\  
             Standard deviation [Mm] & 0.48 & 0.68 & 0.84\\
             Skewness & 1.23 & 1.30 & 1.27\\
             \hline
        \end{tabular}
        \label{table_tr}
    \end{center}
\end{table}    

The average height of the transition region for each run, its standard deviation and skewness are listed in table \ref{table_tr}. The standard deviation increases with resolution for the three different runs, the transition region is therefore not only higher on average but also more corrugated. The skewness has comparable values and is positive for all three runs.

\begin{figure*}
\resizebox{\hsize}{!}{\includegraphics{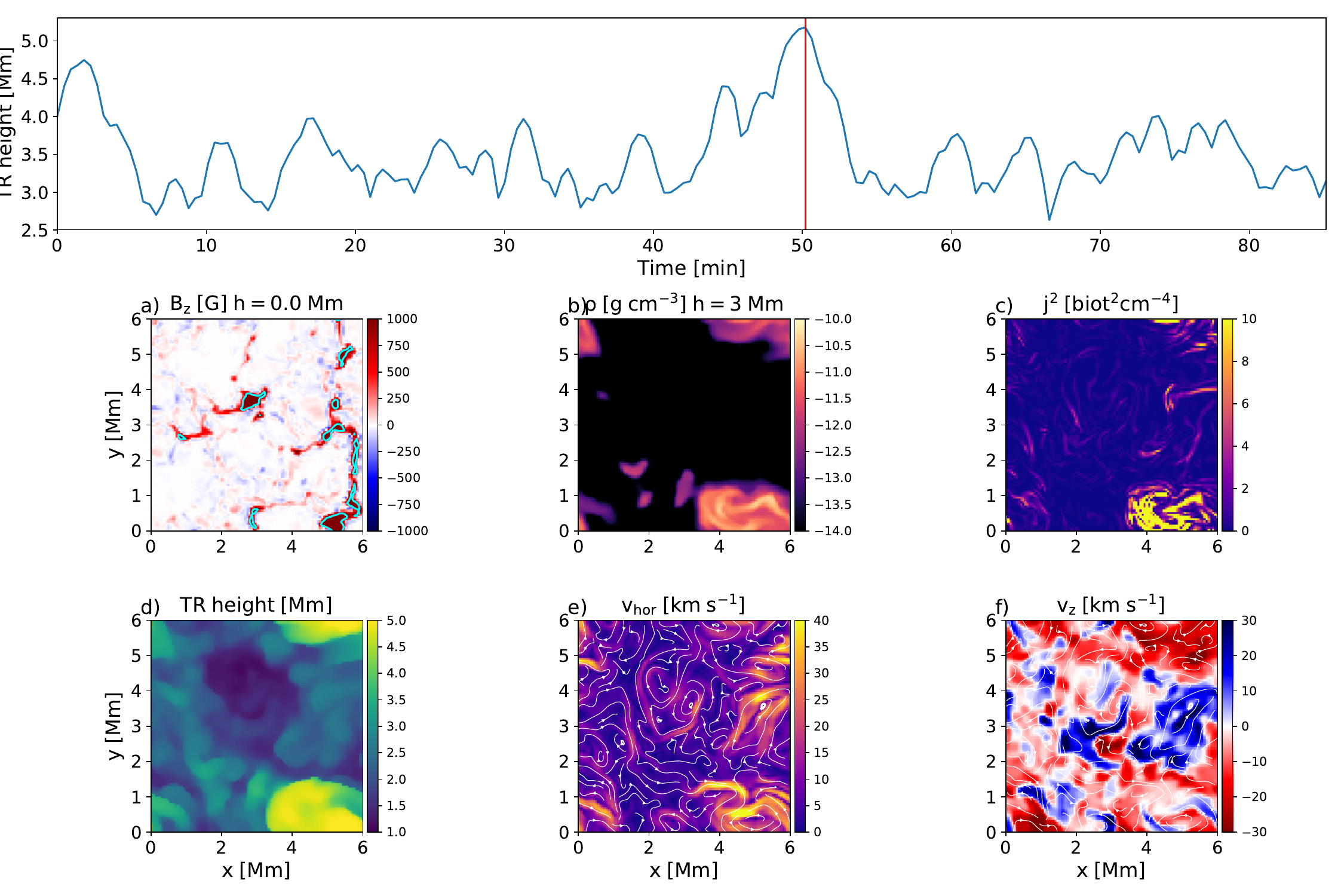}}
    \caption{Highest jet in the LR simulation run at footpoint 2. Top panel: Maximum transition region height as a function of time. Middle: Vertical magnetic field at the photosphere, density and squared current density at a height of 3 Mm above the photosphere, bottom: Transition region height, horizontal velocity and vertical velocity component at a height of 3 Mm above the photosphere. The contours in panel a) outline magnetic concentrations with kilogauss field strength.}
    \label{fig:jet_lr}
\end{figure*}

\begin{figure*}
\resizebox{\hsize}{!}{\includegraphics{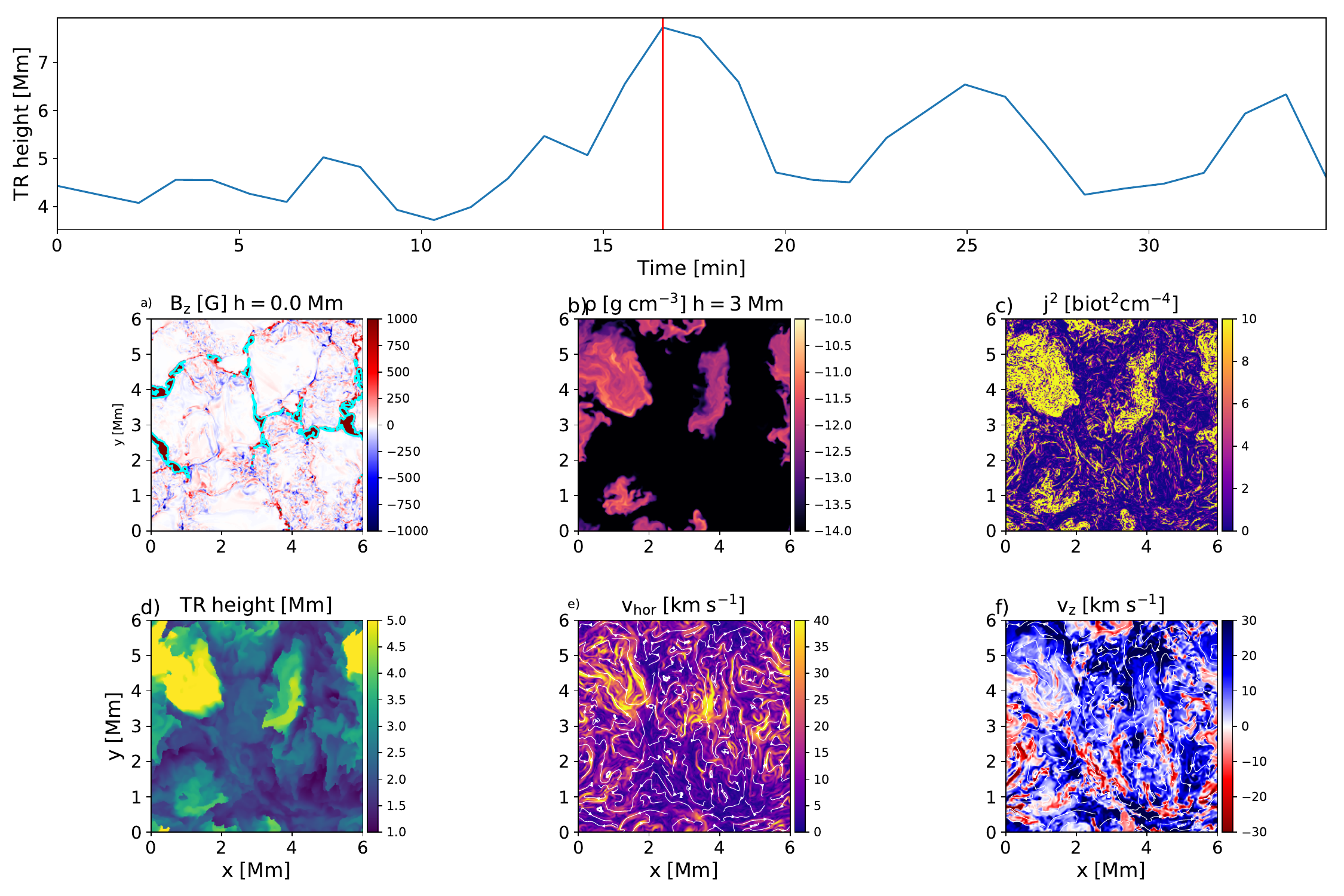}}
    \caption{Highest jet in the HR simulation run at footpoint 1. Top panel: Maximum transition region height as a function of time. Middle: Vertical magnetic field at the photosphere, density and squared current density at a height of 3 Mm above the photosphere, bottom: Transition region height, horizontal velocity and vertical velocity component at a height of 3 Mm above the photosphere. The contours in panel a) outline concentrations with kilogauss field strengths.}
    \label{fig:jet_hr}
\end{figure*}

Comparing the snapshots in which the transition region reaches its highest excursion, we find that while the boundary of the jet has a smooth appearance for the LR run as shown in Fig. \ref{fig:jet_lr}, it is more corrugated for the high-resolution simulation. Density, velocity field and current structures show significantly more complex structures for the HR run (see Fig. \ref{fig:jet_hr}).\\
MURaM uses a five-point-stencil to compute numerical derivatives \citep{2003AN....324..399V}, therefore numerical diffusion affects structures with extents below five grid cells, corresponding to 300 km for the LR run, 120 km for the MR run and 60 km for the HR run. Since the jets have a cross section of a few hundred km at a height of 5 Mm, structures with diameters of a few hundred kilometers will be affected by dissipation in the LR run.\\
The velocity field perpendicular to the loop axis shows more small-scale structure for the HR run. It is also evident from comparing panels (c) in Fig. \ref{fig:jet_lr} and Fig. \ref{fig:jet_hr}, that more current sheets are present at the same height for the HR run, implying a more complex structure of the magnetic field in the chromosphere, transition region and low corona. Higher current densities also lead to higher magnitude of the Lorentz force due to the twisting of field lines that can lift up chromospheric material.\\

\begin{figure}
\includegraphics[width=\columnwidth]{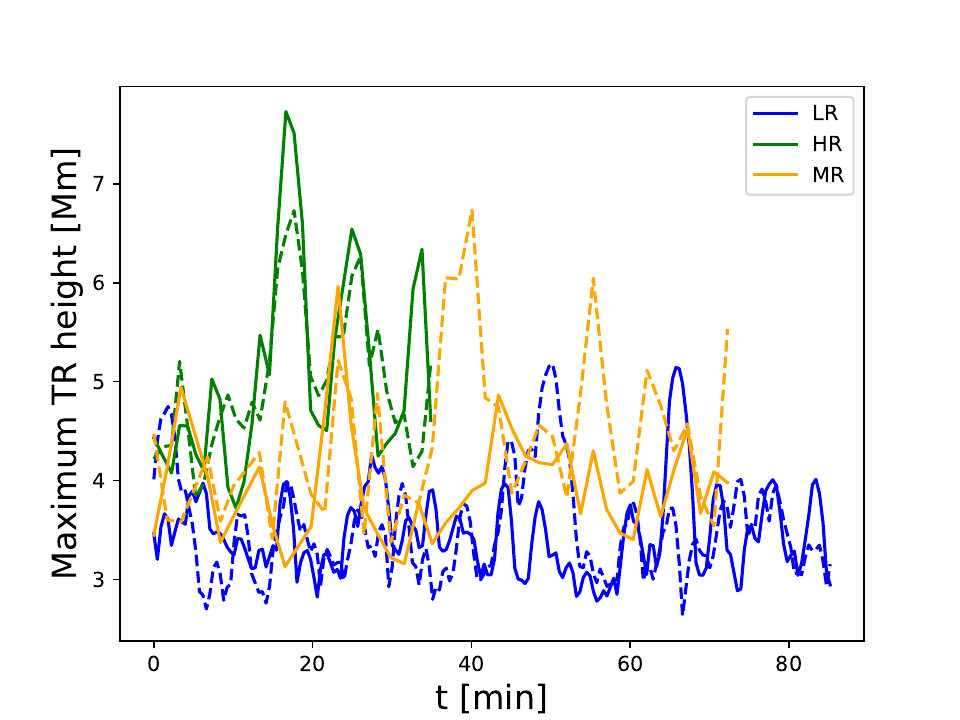}
    \caption{Time evolution of the maximum transition region height. Solid lines show the evolution for footpoint 1 and dashed lines for footpoint 2.}
    \label{fig:tr}
\end{figure}

\section{Discussion}
\label{sec:discussion}

\subsection{Thermodynamic quantities and emission}

Overall, the simulated loops are more dynamic for the MR and HR runs, showing more energetic heating events.
Energy input in the form of upward directed Poynting flux, heating rate, temperature and coronal density increases with a decreasing grid spacing from the LR to the MR run.  Several factors are likely to contribute to this. In the following, we will discuss the energy spectra, the resolution of the steep temperature gradient at the transition region, the photospheric magnetic field, Poynting flux and eventually the effect on the coronal emission.

\subsubsection{Energy spectra}

Both the distribution of energies over spatial scales and bulk properties of the coronal loop change with resolution.
The power spectra for magnetic and kinetic energies as well as the velocity field have excess power at larger scales for the LR and MR simulations. Kinetic and magnetic energy is thus "trapped" at larger scales for low resolution and the grid spacing affects the entire scale distribution of the energy. In addition to the distribution of energies over different spatial scales, the average energy density in the loop midplane is larger for the MR and HR simulation runs. The non-monotonic behaviour of energy densities with resolution could be due to the short runtime of the high resolution simulation since the energy densities vary strongly with time.\\
The energy spectra, especially for the kinetic and internal energy density, are sensitive to the selected time window, but not influenced much by the cadence of snapshots chosen.\\
Magnetic and kinetic energies fluctuate quite strongly over the course of the time series and runs of durations of several hours for each grid size would be necessary to test properly for convergence. This is not feasible for the HR run due to the high computational cost. The MR run is in a state of high magnetic energy due to two sheared flux bundles developing over the course of the run. If the HR simulation was run for longer, it might reach an even higher magnetic energy.

\subsubsection{Comparison to 1D loop models}

The effect of grid spacing on coronal properties has been extensively studied for 1D loop models.
The numerical resolution affects the energy balance across the transition region. The optically thin radiative loss function peaks strongly at transition region temperatures. If the transition region is narrower than the width of one grid cell, optically thin losses will be either under- or overestimated since the emitting volume is too large if the temperature in the grid cell is close to the peak of the radiative loss function, or underestimated if the transition region temperature is not sampled since the narrow transition region would fall between two grid cells.\\ 
The effect of underresolving the transition region on the energy balance is partially mitigated in MURaM by oversampling the transition region in the computation of the optically thin radiative losses (for a detailed description, see \citet{2017_Rempel}). In order to mitigate timestep constraints set by the numerical treatment of heat conduction, MURaM instead uses a hyperbolic diffusion equation for the evolution of the heat flux \citep{2017_Rempel}. This numerical device introduces an artificial limit on the propagation speed of a heat front depending on the maximum wave speed in the simulation, artificially slowing down the heat front. How this numerical treatment affects the enthalpy exchange across the transition region is beyond the scope of this work, but warrants further study.\\
Using 1D hydrodynamic simulations, \citet{Bradshaw_2013} predicted the density to be underestimated by a factor of two and have also found lower coronal temperatures if the transition region is underresolved. Consistent with that, we find both lower temperatures and densities in the LR run compared to the MR and HR runs. The average temperature does not show systematic differences between the MR and HR runs.\\ 
 \citet{Bradshaw_2013} estimated that for a typical active region loop with an apex temperature of 4 MK, a grid spacing of less or equal than 500 m might be needed to resolve the transition region temperature gradient. Even our highest resolution run with a grid spacing of 12 km is very far away from that value. 
Using an adaptive grid, the smallest grid cell width reached in the simulations by \citet{Bradshaw_2013} is 98 m, far out of reach for 3D MHD simulations.
In their numerical experiments, however, \citet{Bradshaw_2013} found that for non-flaring loops with moderate apex temperatures up to 3 MK, a resolution of 25 km is sufficient to model the transition region. This is consistent with our results for thermodynamic quantities in the simulation runs for different grid spacing. While average density and temperature jump between the LR and MR runs, there is less difference between the MR and HR runs. \\
Compared to the simplified 1D loop models by \citet{Bradshaw_2013}, we take into account more factors potentially leading to a decrease of temperature and density in low-resolution simulations. Our simulations are three-dimensional, thus including the tangling of magnetic field lines shuffled around by magnetoconvecion at the footpoints instead of using an ad-hoc heating function. The trends in coronal loop properties are therefore likely not only due to the resolution of the transition region, but also the magnetic field in the photosphere.

\subsubsection{Relation to photospheric magnetic field and Poynting flux}

The photospheric magnetic field is structured on very small scales.
The width of intergranular lanes is on the order of 100 km. In our low-resolution runs with a grid spacing of 60 km the fine structure of the  photospheric magnetic field is not well resolved.\\
We find lower filling factors of kilogauss magnetic fields in the photosphere for the LR run, while the horizontal velocities inside kilogauss magnetic field concentrations do not seem to be affected by resolution, as illustrated in Fig. \ref{fig:fill_fact}. The unsigned photospheric magnetic field is also generally higher for the MR and HR runs.  The dependence of the filling factor of kilogauss magnetic field concentrations on resolution has been studied for the case of a small-scale dynamo in \citet{2014ApJ...789..132R}.  
The behaviour of the turbulent small-scale dynamo is resolution dependent, however, \citet{2014ApJ...789..132R} did not find a systematic resolution dependence of the filling factor in their study. In our case, however, the kilogauss flux concentrations do not arise mainly from the action of a small-scale dynamo, but from the uniform vertical initial magnetic field that is added to the simulation box. We also include coarser resolutions. The lowest resolution in \citet{2014ApJ...789..132R} is 32 km, while our lowest resolution is 60 km, omitting the smallest magnetic structures.\\
The higher coverage of kilogauss magnetic concentrations in the photosphere could partially explain the higher injection of Poynting flux into the solar atmosphere that we find for a smaller grid spacing.\\
The Poynting flux associated with work done by the driver is given by
\begin{equation}
    S_{z}=\frac{1}{4\pi}B_{v}^{2}v_{h}\tan{\theta}, \label{equation:Sz}
\end{equation}
where $B_{v}$ is the vertical magnetic field, $v_{h}$ the horizontal motion of the driver and $\theta$ is the inclination of the magnetic field.  All terms in eq. \ref{equation:Sz} are potentially affected by spatial resolution.
The Poynting flux depends on the magnetic field strength in the photosphere, but also on the horizontal component $B_{h}=B_{z}\tan{\theta}$, so on pre-existing stresses in the solar atmosphere. 
It has been argued that the lower numerical diffusivity in the corona for a smaller grid spacing allows larger gradients to build up in the magnetic and velocity field, allowing more Poynting flux to enter the corona due to the increased inclination of the magnetic field lines. Likewise, more free energy can build up in the corona before being dissipated in heating events \citep{2015_Klimchuk}. \citet{2008ApJ...677.1348R} found that quantitative and qualitative aspects of heating were dependent on the Reynolds number. It is not sufficient to estimate the energy input from the driver at the solar surface to estimate the energy injection into the corona as the adjustment of the magnetic field configuration in the corona to the injected energy will have a back-reaction on the energy input. 
Because the diffusivity in MURaM is resolution dependent, we would therefore expect that the numerical resolution increases the Poynting flux able to enter the corona.\\
\citet{2019_Martinez_Sykora} report that the component of the Poynting flux related to vertical transport of horizontal magnetic field in the chromosphere is negative up to 0.2 Mm and positive above 0.75 Mm, indicating that magnetic field is advected into the corona above this height. 
We find a different behaviour of this component of the chromospheric Poynting flux in our study, which is likely caused by the differences in setup.
The horizontal field transport component of the Poynting flux is negative throughout, and above the photosphere the total Poynting flux is dominated by the action of horizontal motions on almost vertical field (field-shaking component). 
While \citet{2019_Martinez_Sykora} start their simulation with a weak vertical seed field of 2.5 G that is amplified by convective motions, our simulation is the simulation is started with a relatively strong vertical magnetic field of 60 G. This likely causes the contribution of horizontal flows acting on almost vertical magnetic field to dominate the total Poynting flux.\\

While the current density and thus small-scale gradients in the magnetic field in our numerical experiments increase monotonically with decreasing grid spacing, the picture is not so clear for the large-scale structure of the magnetic field due to the strong fluctuation of the perpendicular component of the magnetic field in time. Longer time series would be needed to determine whether on average the coronal magnetic field is more inclined on large scales in the MR and HR simulations.
 Higher numerical resolution leads to a stronger photospheric magnetic field and therefore to a larger energy reservoir that can contribute to chromospheric and coronal heating.
From the elevated Poynting flux especially in the chromosphere we can conclude that more energy is pumped into the high-resolution loop from the photosphere in the first place.

\subsubsection{Consequences for amount and structuring of coronal emission}

The higher temperatures and increased density for the MR and HR runs lead to higher emission with smaller grid spacing in all three of the AIA channels we examined. Similar to the trend in average coronal temperatures and densities, however, the coronal emission seems to saturate from a grid spacing of 24 km onward.\\
We also compared the relative emission in different channels. The fact that the emission in all three AIA channels increases with resolution, but the ratio of count rates in the 193 \AA\; and 211 \AA\; channel to the count rates in the 171 \AA\; channel stays roughly the same for different resolutions indicates that the density changes with resolution have a larger influence than the only moderately increased temperature. The lower densities in the LR run could to some degree counteract the effect of lower energy input since the heating rate per particle will be higher for lower densities. \\
The increased tangling of the magnetic field leads to stronger current sheets, higher Lorentz forces and, as a consequence, higher plasma velocities. This is consistent with the trend for the current density and the Lorentz force found in \citet{2012ApJ...747..109N} for different explicit resistivities. The higher perpendicular velocities that we find for higher resolutions will lead to increased non-thermal line width for synthetic line profiles \citep{2024a_Breu}. The average Doppler velocities measured using a line of sight perpendicular to the loop axis, however, do not change significantly with resolution. Increased numerical resolution thus leads to stronger small-scale flows, but not necessarily to stronger large-scale flows. 
In addition of the magnitude of the emission, sizes of bright structures in the corona are affected by the grid spacing.
Structuring of the emission is present at all scales in the corona.
The distribution of widths of coronal loop strands peaks at smaller widths for high resolution, making it difficult to determine a "fundamental" strand size from the simulations in this study. Even for the higher resolution runs, however, structures with scales on the order of Mm are still present, with small-scale spatial fluctuations superimposed on the large-scale structures. This finding remains the same for both methods we used to identify structure sizes and is consistent with more energy being present in larger structures for low-resolution simulations. The clear peak in the size distributions vanishes if the synthetic observations are degraded to the resolution of the AIA instrument, indicating that instruments with better spatial resolution might find smaller strands.\\
Doppler shifts and line widths as would be observed with MUSE have been investigated in \citet{2024_Cozzo} and \citet{2024_Breu_b}.  Both simulations show possible signatures in the form of alternating blue- and redshifts in coronal lines. While rotational motions occur at least initially on relatively large scales of about 2 Mm in the MHD avalanche simulation by \citet{2024_Cozzo}, the size distribution of vortices reaches all the way down to a few 10s of km in the HR run, a scale below the resolving capabilities of MUSE. The superposition of many small-scale vortices along the LOS would also make it difficult to discern individual structures. The nonthermal linewidths seen in the simulation by \citet{2024_Cozzo} are larger than the line widths even in the HR run. This might be due to the more violent outflows triggered by the kink instability.\\
 While some effects of low resolution, such as not accurately modelling the energy transfer through the transition region, could be mitigated (for example using the transition region adaptive conduction (TRAC) method \citep{2020_Johnston, 2021_Johnston}), other effects of lack of resolution, such as the lower energy input at the photosphere and the small-scale structuring of coronal emission,  would still be present.\\
It is important to note that the temporal variation of averaged quantities in the corona is quite large for all runs and due to the computational cost only a short time series is available for the highest resolution. If only a subset of the time range is chosen for averaging, the separation between the average quantities for the different runs would change considerably. The shaded areas illustrating the standard deviations of the temporal variations of the respective quantities depicted in Fig. \ref{fig:quant_evo} overlap for all three runs in case of the temperature and heating rate as well as the vertical velocity and for the MR and HR runs for most quantities apart from the current density and the perpendicular component of the magnetic field. To quantify better the trends of these quantities with resolution, longer time series would be necessary.
Overall, we find that the MR and HR runs yield rather similar results for most quantities apart from the current density and the Poynting flux in the chromosphere, despite the HR simulation including more fine structure. A spatial resolution of about 24 km should therefore be sufficient for most purposes.\\

\subsection{Dynamics of jets at the loop base}

Chromospheric jets are more ubiquitous, occur more frequently, are more energetic and reach larger heights for the MR and HR runs. The jets have a more complex structure, likely due to the smaller scales in the magnetic and velocity fields. This is consistent with the generally higher energy input into the corona for the MR and HR simulations. The jets could also be an important contributor to the non-thermal line width of spectral lines near the loop footpoints, which increases with increasing resolution \citep{2024a_Breu}. Choosing a high enough resolution is therefore important for correctly modelling the dynamic behaviour of the lower atmosphere.\\
The height of chromospheric jets is dependent on the coronal temperature, as found in a study by \citet{2015_Iijima}. In our case, we find higher jets in the MR and HR run, despite the average temperature being slightly higher than in the LR run. It seems that the effect of the grid size on the jet height dominates over the influence of the slightly higher temperature. It also has to be noted that the temperature differences considered in \citet{2015_Iijima} are much larger than the relatively small average differences between our runs.
It is important to point out that the simulations described in this paper assume local thermodynamic equilibrium (LTE) radiative transfer in the chromosphere. The properties of the jets therefore might be affected if effects from non-local thermodynamic equilibrium, which are important in the chromosphere, are included in the simulations. Non-LTE effects have recently been added to the MURaM code (see \citet{2022_Przybylski}) based on the seminal work by \citet{2011_Leenaarts}. 
While these jets show a spicule-like appearance, they do not reach the velocities observed in type 2 spicules.

\section{Conclusions}
\label{sec:conclusions}

The aim of this study is to quantify the effect of resolution on observations synthesized from MURaM data cubes, considering both the average coronal emission as well as its structure due to strand- and jet-like features. In general, higher numerical grid resolutions (i.e.~resolving smaller spatial scales) leads to a more complex and dynamic solar atmosphere.\\ 
A common argument to justify low resolution in large-scale simulations is that the simulation will develop an equilibrium in which the dissipated energy will eventually balance the energy injected into the system via the Poynting flux. However, such models do not reproduce the small-scale explosive nature of the heating and while they reproduce large-scale structures in solar observations fairly well, the underlying heating process might be qualitatively very different from the real Sun.\\
Every part of the coupled convection zone-chromosphere-transition region-corona system is affected by resolution, from the energy generation by magnetoconvection to energy and mass transport and conversion into thermal energy, making it difficult to disentangle cause and effect for changes of the behaviour of the simulations with resolution.
Energy injection in the form of Poynting flux, temperature and density increase initially with resolution between the LR and MR run. Some quantities seem to saturate for even higher resolution, while others do not show signs of saturation for the resolutions tested in this study. While the average coronal temperature is only moderately affected, current density and the velocity components perpendicular to the loop axis have a stronger dependence on the numerical resolution. While smaller grid size leads to stronger current sheets, they tend to be more localized and fragmented, as visible in Fig. \ref{fig:ap_cut_difres}. Stronger current sheets lead to larger Lorentz forces accelerating plasma and therefore to more violent heating events. The larger velocities that are reached in the MR and HR runs lead to larger non-thermal widths of spectral lines.\\
The resolution not only influences thermodynamic quantities, but also the stratification of the loop. The transition region reaches larger heights with increasing resolution. Choosing the resolution carefully is therefore not only important for correctly modelling the shape of emission lines which are affected by the velocity distribution, but also the height of chromospheric jets.\\
Since the optically thin emission depends roughly on the square of the density, it is strongly affected by a change in resolution.
Therefore, synthetic observables are affected despite the moderate dependence of loop temperature on resolution. The emission will be underestimated if the grid resolution is chosen too coarse. However, the ratio between the loop-integrated emission in different passbands does not strongly depend on resolution.\\
For simulations at all resolutions, coronal strands exist on a range of scales. While strands in the emission synthesized from the high-resolution run have more structure at fine scales, the emission is still organised into structures with larger widths. The strand widths, however, peak at different values for different resolutions.
In order to test whether spatially and temporally averaged coronal quantities converge with increasing resolution, time ranges of several hours of solar time would be required. The computational cost especially for the high resolution simulations is (currently still) prohibitive.\\
While an increasing amount of fine structure is present for smaller and smaller grid spacings and some quantities, such as the current density, show no signs of converging, average temperatures, coronal emission and velocities do not strongly increase between the medium and high resolution runs. 
To conclude, the study presented here shows that a grid spacing of about 24 km should be adequate for comparing
simulation results with data from current and upcoming missions, such as Solar Orbiter EUI and MUSE.

\section*{Acknowledgements}

This project has received funding from the European
Research Council (ERC) under the European Union’s Horizon 2020 research
and innovation program (grant agreement No. 101097844). We gratefully acknowledge the computational resources provided by the Cobra supercomputer system of the Max Planck Computing and Data Facility (MPCDF) in Garching,
Germany. 
The research leading to these results has received funding from the UK Science and Technology Facilities Council (consolidated grant ST/W001195/1).
Ineke De Moortel received funding from the Research Council of Norway through its Centres of Excellence scheme, project number 262622.
CHIANTI is a collaborative project involving George Mason University, the University of Michigan (USA), University of Cambridge (UK) and NASA Goddard Space Flight Center (USA).

%%%%%%%%%%%%%%%%%%%%%%%%%%%%%%%%%%%%%%%%%%%%%%%%%%
\section*{Data Availability}

Due to their size, the data from the numerical simulations and analysis
presented in this paper are available from the corresponding author
upon request.



\bibliographystyle{mnras}
\bibliography{main.bib} % if your bibtex file is called example.bib

%\begin{thebibliography}{99}
%\end{thebibliography}

%%%%%%%%%%%%%%%%%%%%%%%%%%%%%%%%%%%%%%%%%%%%%%%%%%

%%%%%%%%%%%%%%%%% APPENDICES %%%%%%%%%%%%%%%%%%%%%

\begin{appendix}

\section{Effective resistive and viscous diffusivities}
We do not use an explicit resistivity or viscosity in MURaM. Instead, the finite  numerical resolution leads to diffusion of magnetic and velocity fields. 
The heating rate due to finite resistivity for a uniform value of the magnetic diffusivity is
\begin{equation}
    Q_{res}=\frac{\eta}{4\pi}(\nabla \times \mathbf{B})^{2}=4\pi \eta \mathbf{j}^{2}.
\end{equation}
We do not specify a uniform magnetic diffusivity. Instead, viscous and resistive heating are due to numerical resistivity. In order to prevent the buildup of energy on the grid scale, MURaM uses a slope-limited diffusion scheme \citep{2014ApJ...789..132R}. 
Magnetic and viscous diffusivities are not uniform in the domain, but instead enhanced at the location of gradients in the magnetic and velocity field. Following \citet{2018_Rempel}, effective resistive and viscous diffusivities can be estimated as
\begin{align}
    \nu_{\rm{eff}} &= \frac{\langle Q_{vis,num}\rangle}{\left\langle\rho\sum_{i,k}\frac{\delta v_{i}}{\delta x_{k}}\left [\frac{\delta v_{i}}{\delta x_{k}}+\frac{\delta v_{k}}{\delta x_{i}} - \frac{2}{3}\delta_{ik}\nabla \cdot \mathbf{v}\right ]\right\rangle},\\
    \eta_{\rm{eff}} &= 4\pi \frac{\langle Q_{res,num}\rangle}{\langle |\nabla \times \mathbf{B}|^{2}\rangle}.
\end{align}
Here $\langle Q_{vis,num}\rangle$ and $\langle Q_{res,num}\rangle$ are resistive and viscous numerical heating rates averaged over the coronal volume. Effective magnetic and viscous diffusivities are dependent on resolution. We have listed effective values for the diffusivities as well as the effective Prandtl number for the coronal part of the simulation domain in Table \ref{table1}. 
\begin{table*}

\begin{center}
\caption{Effective diffusivities and magnetic Prandtl number for different grid sizes.}
    \begin{tabular}{l |l |l |l |l }
          & 60 km & 24 km & 12 km & Spitzer value \\
          \hline
         $\eta_{eff}\; \rm{[cm^{2}s^{-1}]}$ & $1.8\times 10^{11}$ & $4.5\times 10^{10}$ & $2.2\times 10^{10}$& 2631 \\
         \hline
         $\nu_{eff}\; \rm{[cm^{2}s^{-1}]}$ & $3.1\times 10^{13}$ & $6.9\times 10^{12}$ & $1.7\times 10^{12}$ & $6.9\times 10^{14}$\\
         \hline
         $P_{\rm{m,eff}}$ & 172.2 & 153.3 & 77.2 & $2.6\times 10^{11}$
    \end{tabular}
    \label{table1}    
\end{center}
\end{table*}

%The diffusivities were computed following \citet{2018ApJ...859..161R}. 
% Due to the five-point stencil used in MURaM, for run LR structures up to sizes of 300 km will be affected by diffusion.
For the setup used in this study, increasing the resolution by a factor of five reduced
the effective magnetic diffusivity by roughly a factor of eight and the effective viscous
diffusivity by a factor of 18. In a realistic Prandtl number regime for the Sun, the resistive diffusivity
would be negligible compared to the viscous diffusivity.
Numerical resistive and viscous heating rates are strongly intermittent in space and time
and not necessarily well-correlated with current structures and strain rate for a single
timestep \citep{2017_Rempel}. Therefore, effective viscous and resistive diffusivities are not well defined.\\
One has to be careful in comparing effective Prandtl numbers for different resolutions since the diffusivity of the code for different variables is influenced by many factors, including the Alfv\'{e}n speed limiter.

\end{appendix}

%%%%%%%%%%%%%%%%%%%%%%%%%%%%%%%%%%%%%%%%%%%%%%%%%%

% Don't change these lines
\bsp	% typesetting comment
\label{lastpage}
\end{document}